\documentclass[reqno,a4paper,final,11pt]{amsart}
\usepackage[british]{babel}
\usepackage{amssymb,amstext,amsthm,eucal}
\usepackage{bbold}
\usepackage{array,color}
\usepackage[utf8]{inputenc}
\usepackage[notref,notcite]{showkeys}
\usepackage[small]{eulervm}
\usepackage{tgpagella}
\usepackage{hyperref}
\hypersetup{%
  pdftitle   = {The homogeneity theorem for supergravity backgrounds},
  pdfkeywords = {supergravity, Killing spinors, isometries,
    homogeneity, supersymmetry},
  pdfauthor  = {José Figueroa-O'Farrill},
  pdfcreator = {\LaTeX\ with package \flqq hyperref\frqq}
}
\newcommand{\half}{\tfrac{1}{2}}

\newcommand{\fk}{\mathfrak{k}}

\newcommand{\SO}{\mathrm{SO}}

\newcommand{\Cl}{\mathrm{C}\ell}
\newcommand{\Spin}{\mathrm{Spin}}
\ifx\Sp\UnDeFiNeD
\newcommand{\Sp}{\mathrm{Sp}}
\else
\renewcommand{\Sp}{\mathrm{Sp}}
\fi
\renewcommand{\SS}{\mathbb{S}}

\newcommand{\RR}{\mathbb{R}}

\DeclareMathOperator{\im}{im}

\newcommand{\1}{\mathbb{1}}
\newcommand{\spb}{\$}

\definecolor{orange}{rgb}{0.9,0.45,0}
\newcommand{\MUNCH}[1]{\relax}

\allowdisplaybreaks[1]
\begin{document}
\title[Homogeneity of supergravity backgrounds]{The homogeneity theorem for supergravity backgrounds}
\author[Figueroa-O'Farrill]{José Figueroa-O'Farrill}
\author[Hustler]{Noel Hustler}
\address{Maxwell and Tait Institutes, School of Mathematics, University of Edinburgh}
\thanks{EMPG-12-14}
\begin{abstract}
  We prove the strong homogeneity conjecture for eleven- and ten-dimensional (Poincaré) supergravity backgrounds.  In other words, we show that any backgrounds of 11-dimensional, type I/heterotic or type II supergravity theories preserving a fraction $\nu > \frac12$ of the supersymmetry of the underlying theory are necessarily locally homogeneous.  Moreover we show that the homogeneity is due precisely to the supersymmetry, so that at every point of the spacetime one can find a frame for the tangent space made out of Killing vectors constructed out of the Killing spinors.
\end{abstract}
\maketitle
\tableofcontents

\section{Introduction}

It is a fact that all known (Poincaré) supergravity backgrounds in 10 and 11 dimensions (and also in some lower dimensions) which preserve more than one half of the supersymmetry of the theory are homogeneous, by which one means that there is a Lie group acting transitively on the spacetime and preserving all the bosonic fields which are turned on in the background: metric, fluxes,...  This empirical fact led naturally to the homogeneity conjecture, reviewed in \cite{JMF-HC-Lecs}, which we heard for the first time in a private communication from Patrick Meessen in 2004.  The earliest attempt to prove this conjecture was \cite{FMPHom}, where it was shown that classical M-theory backgrounds preserving more than $\frac34$ of the supersymmetry are (locally) homogeneous.  As explained in \cite{FMPHom}, we usually work with local metrics defined on open subsets of $\RR^d$, whence the relevant notion is that of \emph{local} homogeneity.  This follows from the related notion of local transitivity, which simply says that at every point in the spacetime there is a frame made out of Killing vectors which preserve all bosonic fields.  If we further demand that the frame consists of Killing vectors which are made out of the Killing spinors of the background we arrive at what one could call the ``strong'' form of the conjecture.  In \cite{FMPHom} there is a proof of a strong form of the conjecture but for backgrounds preserving more than $\frac34$ of the supersymmetry and moreover suggested that the critical supersymmetry fraction $\nu_c$ beyond which homogeneity was guaranteed was in fact $\frac34$.  That suggestion was based on what turns out to be an error in that paper, namely the construction of a 24-dimensional subspace of the spinor representation of $\Spin(1,10)$ obeying certain properties.  In fact, as will be explained below, no such subspaces exist.

In \cite{EHJGMHom} it was shown that any type IIB supergravity background preserving more than $\frac34$ of the supersymmetry is (locally) homogeneous and moreover that the homogeneity is due to the supersymmetry.  In the same paper, the conjecture (for  $\nu_c = \half$) was proven in type I/heterotic supergravity, albeit not in the strong form.  In other words, the question was left open whether the homogeneity for type I/heterotic backgrounds preserving more than $\half$ of the supersymmetry could be ``accidental'', as the proof used unrelated results concerning parallelisable heterotic backgrounds.

The purpose of this note is to prove the strong form of the conjecture for eleven-dimensional, type I/heterotic and type II supergravity backgrounds.  The proof is quite elementary and uniform in all cases, resting as it does on two fundamental facts.  Firstly, that there is a ``squaring'' map from spinor fields to vector fields such that when applied to Killing spinors produces Killing vectors which in addition preserve all the other bosonic fields which are turned on in the background.  This has been established in \cite{GauPak} (but see also \cite{FMPHom}) for eleven-dimensional supergravity and in \cite{EHJGMHom} for type IIB and type I/heterotic supergravities.  And lastly, that this squaring map is pointwise surjective provided that the dimension of the space of Killing spinors is greater than one half of the rank of the spinor bundle.  It is the proof of this latter fact which is the main aim of this note.

The note is organised as follows.  In Section \ref{sec:general-set-up} we present the general set-up, since the idea of the proof is to a large extent independent of the details of the supergravity theory.  Then in Section \ref{sec:proof-homog-conj} we present the proofs for each of the supergravity theories under discussion.  Finally we conclude in Section~\ref{sec:conclusion}.

\section{General set-up}
\label{sec:general-set-up}

Unless otherwise stated, by ``supergravity background'' we shall mean a bosonic background of eleven-dimensional, type II or type I/heterotic supergravity.  The common ingredients in all supergravity backgrounds are a lorentzian spin manifold $(M,g)$ and a bundle $\spb \to M$ of spinors on which one has defined a connection $D$.  The connection $D$ depends on the bosonic fields of the background in question.  The bundle $\spb$ is obtained from the spin bundle as a vector bundle associated to a representation $S$ of the spin group.  The representation $S$ need not be irreducible, but might be a direct sum of irreducible spinor representations.  The tangent bundle $TM$ is similarly obtained, but this time the associated representation of the spin group is the vector representation of the corresponding orthogonal group and shall be denoted $V$.

Each supergravity background has a notion of Killing spinor, which is a section of $\spb$ which is $D$-parallel and in some cases might satisfy additional algebraic equations which say that it is in the kernel of some bundle maps $\spb \to \spb$ which depend on the bosonic fields in the supergravity background.  Since the equations defining a Killing spinor are linear, the Killing spinors span a vector space which we denote $\fk_1$.  Because the equations satisfied by Killing spinors are at most first order in derivatives, a Killing spinor is uniquely determined by its value at any point $p \in M$, whence having chosen such a point, $\fk_1$ can be identified with a subspace of the fibre $\spb_p$ of $\spb$ at $p$, which can itself be identified with the representation $S$.  This means that we can think of $\fk_1$ as a vector subspace of $S$.

Another common ingredient of supergravity backgrounds is the existence of a symmetric bilinear bundle map $\Phi: \spb \times \spb \to TM$ (called ``squaring'') with the property that if $\varepsilon_{1,2}$ are Killing spinors, then $\Phi(\varepsilon_1,\varepsilon_2)$ is a Killing vector which moreover preserves all the other bosonic fields in the background (when appropriate, only up to gauge transformations). Let $\fk_0 = \Phi(\fk_1,\fk_1)$ denote the vector fields obtained by squaring Killing spinors.  Then it was shown in \cite{FMPHom} and \cite{EHJGMHom} for eleven- and ten-dimensional supergravity backgrounds, respectively, that on the 2-graded vector space $\fk = \fk_0 \oplus \fk_1$ one can define the structure of a Lie superalgebra, called the \emph{Killing superalgebra} of the supergravity background.  In particular, $\fk_0$ is a Lie algebra and the strong homogeneity conjecture says that $\fk_0$ acts locally transitively on the spacetime; that is, that the values at $p$ of the Killing vectors in $\fk_0$ span $T_pM$ for all $p \in M$.

Fixing a point $p \in M$ once and for all and identifying $\spb_p$ with with $S$, the squaring map $\Phi$ induces a spin-equivariant symmetric bilinear map $\varphi: S \times S \to V$.  Being symmetric and bilinear, $\varphi$ is uniquely determined by its value on the diagonal by the usual polarisation identity
\begin{equation}
  \label{eq:polarisation}
  2\varphi(\varepsilon_1,\varepsilon_2) = \varphi(\varepsilon_1+\varepsilon_2,\varepsilon_1+\varepsilon_2) - \varphi(\varepsilon_1,\varepsilon_1) - \varphi(\varepsilon_2,\varepsilon_2)~.
\end{equation}
A final property of the map $\varphi$ is that for any $\varepsilon \in S$, the vector $\varphi(\varepsilon,\varepsilon)$ is either timelike or null relative to the lorentzian inner product on $V$ induced by the restriction to $T_pM$ of the metric $g$.  Of course, for two different $\varepsilon_{1,2} \in S$, the causal type of $\varphi(\varepsilon_1,\varepsilon_2)$ is typically unrestricted.

The proof of the conjecture consists in showing that if $\dim \fk_1 > \half \dim S$, then the restriction $\varphi|_{\fk_1} : \fk_1 \times \fk_1 \to V$ of the map $\varphi$ to $\fk_1$ is surjective onto $V$.  In other words, that we can always find a frame for $T_pM$ which is made out of the values at $p$ of Killing vectors in the image $\Phi(\fk_1,\fk_1)$, thus proving the strong form of the homogeneity conjecture.

The idea of the proof is the same in all cases.  To show that $\varphi|_{\fk_1}$ is surjective one would like to show that the perpendicular complement of its image in $V$ is trivial.  It is not difficult to show that in all cases $(\im \varphi|_{\fk_1})^\perp$ is a totally null subspace of $V$, whence in lorentzian signature its dimension is bounded above by 1.  One concludes the proof by showing that the case of where the dimension is 1 cannot occur by deriving a contradiction.

\section{The proof of the strong homogeneity conjecture}
\label{sec:proof-homog-conj}

For each of the supergravity theories in question we will now exhibit the map $\varphi: S \times S \to V$, which requires in particular identifying the representations $S$ and $V$, and we will prove that the restriction of $\varphi$ to any subspace $W \subset S$ of dimension $\dim W > \half \dim S$ is surjective.

\subsection{Eleven-dimensional supergravity}
\label{sec:elev-dimens-supergr}

In eleven-dimensional supergravity, $S$ is one of the two irreducible representations of the Clifford algebra $\Cl(1,10)$ and hence restricts to $\Spin(1,10)$ as its unique irreducible spinor representation.  It is real and 32-dimensional and admits an invariant symplectic structure we shall denote $\left(-,-\right)$.  $V$ is the eleven-dimensional real vector representation of $\SO(1,10)$ and we shall let $\left<-,-\right>$ denote the invariant lorentzian inner product.  The squaring map $\varphi : S \times S \to V$ is the transpose of the Clifford action $V \times S \to S$ relative to the symplectic structure on $S$ and the lorentzian inner product on $V$.  In other words, for every $v \in V$ and $\varepsilon_{1,2} \in S$,
\begin{equation}
  \left<v, \varphi(\varepsilon_1,\varepsilon_2)\right> = \left(\varepsilon_1, v \cdot \varepsilon_2\right)~.
\end{equation}
Choosing a pseudo-orthonormal basis $(e_\mu)$ for $V$ and letting $(\Gamma_\mu)$ denote the corresponding gamma matrices, we have
\begin{equation}
  \varphi(\varepsilon_1,\varepsilon_2) = \bar\varepsilon_1 \Gamma^\mu \varepsilon_2 e_\mu~,
\end{equation}
where we have used the standard physics notation $\bar\varepsilon_1 \varepsilon_2$ for the symplectic inner product on $S$, which is defined as
\begin{equation}
  \bar\varepsilon_1 \varepsilon_2 = \varepsilon_1^\dagger \Gamma_0 \varepsilon_2~.
\end{equation}
By taking $\varepsilon_1 = \varepsilon_2 = \varepsilon$, this shows that the vector
\begin{equation}
    v_\mu := \bar\varepsilon \Gamma_\mu \varepsilon = \varepsilon^\dagger \Gamma_0 \Gamma_\mu \varepsilon
\end{equation}
obtained by squaring $\varepsilon \neq 0$ has a nonzero component along $e_0$:
\begin{equation}
  v_0 = \bar\varepsilon \Gamma_0 \varepsilon = \varepsilon^\dagger \Gamma_0 \Gamma_0 \varepsilon = - \varepsilon^\dagger \varepsilon = - |\varepsilon|^2 < 0~.
\end{equation}
This means that $v$ is either null or timelike, since if it were spacelike one could Lorentz-transform to the rest frame, where $v_0 = 0$.

The values of Killing spinors at the point $p$ span a subspace $W \subset S$.  We want to show that if $\dim W > 16$  then the restriction of $\varphi$ to $W$ is surjective onto $V$.

Let $\dim W > 16$.  The map
\begin{equation}
  \varphi|_W : W \otimes W \to V
\end{equation}
is surjective if and only if the perpendicular complement of its image is trivial.  Equivalently, if and only if the only vector $v \in V$ obeying
\begin{equation}
  \label{eq:condition}
  (\varepsilon_1, v \cdot \varepsilon_2) = 0\qquad \text{for all $\varepsilon_{1,2} \in W$}
\end{equation}
is the zero vector $v=0$.

Our first observation is that any $v\in V$ satisfying \eqref{eq:condition} is necessarily null.  Indeed, notice that \eqref{eq:condition} can be rephrased as saying that the Clifford product by $v$ sends $W \to W^\perp$, where
\begin{equation}
  W^\perp = \left\{ \varepsilon \in S \mid (\varepsilon,w) = 0 \quad
    \text{for all $w\in W$} \right\}
\end{equation}
is the symplectic perpendicular complement of $W$.  Since $\dim W  + \dim W^\perp  = 32$, it follows from $\dim W > 16$ that $\dim W^\perp < 16$, whence the Clifford product by $v$ must have nonzero kernel purely on dimensional grounds.  On the other hand, $v^2 = -|v|^2 \1$, whence $v$ has nonzero kernel if and only if $|v|^2 = 0$.  (Here $|v|^2 = \left<v,v\right>$ denotes the indefinite norm in $V$.)

In other words, the perpendicular complement (relative to $\left<-,-\right>$) of the image of $\varphi|_W$ is a totally null subspace of $V$.  Since $V$ is lorentzian, any totally null subspace is at most one-dimensional.  Moreover, if one-dimensional, it is spanned by a null vector in $V$.

Assume for a contradiction that the perpendicular complement of the image of $\varphi|_W$ is one-dimensional.
Without loss of generality we can choose a Witt basis $(e_+,e_-,e_i)$ for $V$ such that the image of $\varphi|_W$ is the perpendicular complement of $e_+$, which is spanned by $e_+$ itself and the $e_i$.  In particular, this means that for every $\varepsilon \in W$, $\varphi(\varepsilon,\varepsilon)$ must be perpendicular to $e_+$, but we have seen that it cannot be spacelike, hence it has to be collinear with $e_+$.  In other words, for any $\varepsilon\in W$, $\varphi(\varepsilon,\varepsilon) = \lambda(\varepsilon) e_+$, for some function $\lambda: W \to \RR$.  Now consider $\varphi(\varepsilon_1,\varepsilon_2)$ for $\varepsilon_1,\varepsilon_2 \in W$.  By the polarisation identity \eqref{eq:polarisation},
\begin{equation}
  \begin{split}
    2 \varphi(\varepsilon_1,\varepsilon_2) &= \varphi(\varepsilon_1+\varepsilon_2,\varepsilon_1+\varepsilon_2) - \varphi(\varepsilon_1,\varepsilon_1) - \varphi(\varepsilon_2,\varepsilon_2)\\
    &= \lambda(\varepsilon_1+\varepsilon_2) e_+ - \lambda(\varepsilon_1) e_+ - \lambda(\varepsilon_2) e_+\\
    &= \left(\lambda(\varepsilon_1+\varepsilon_2) - \lambda(\varepsilon_1) - \lambda(\varepsilon_2) \right) e_+~,
  \end{split}
\end{equation}
whence the image of $\varphi|_W$ is contained in the null line spanned by $e_+$, which contradicts the fact that its perpendicular complement is one-dimensional.  This means that $\varphi|_W$ has to be surjective, as desired.

In \cite[§6.3]{FMPHom} it was claimed that there was a 24-dimensional subspace $W \subset S$ with the property that the image of $\varphi|_W$ was not all of $V$.  The ``proof'' of this statement in that paper is incorrect.  It relies on choosing the signature of the bilinear form $\beta(\varepsilon_1,\varepsilon_2) := \left(\varepsilon_1,\Gamma_+ \varepsilon_2\right)$ which is not possible, as the signature of this bilinear form is not a matter of choice but follows from a calculation.  Indeed, one can compute it and it has rank 16 and it is semi-definite.

\subsection{Type IIA supergravity}
\label{sec:type-iia-supergr}

We may prove the strong homogeneity conjecture for type IIA supergravity as a consequence of the one for eleven-dimensional supergravity.  Indeed, any background of IIA supergravity preserving more than half of the supersymmetry oxidises to a background of eleven-dimensional supergravity which also preserves more than half of the supersymmetry.  By the above result, it is locally homogeneous.  The eleven-dimensional geometry is the total space of a locally trivial fibre bundle over the IIA geometry.  The Killing spinors of the eleven-dimensional supergravity background obtained via oxidation are constant along the fibres, meaning that their Lie derivative along the Killing vector along the fibres vanishes.  This is in fact that geometric interpretation of the vanishing of the supersymmetry variation of the dilatino.  This means that the Killing vectors obtained by squaring them are also constant along the fibres, which means that they commute with the Killing vector along the fibre and hence push down to Killing vectors of the IIA background.  Since they act locally transitive in the eleven-dimensional geometry, their push-downs to the base also act locally transitively.  This shows that the IIA background is locally homogeneous.

\subsection{Type I/Heterotic supergravity}
\label{sec:type-ihet-supergr}

In type I/heterotic supergravity the relevant spinor representation is $S_+$, the positive-chirality spinor representation of $\Spin(1,9)$ which is real and 16-dimensional.  We will let $S = S_+ \oplus S_-$ be the unique irreducible Clifford module of $\Cl(1,9)$.  It is $S$ which has an invariant symplectic inner product $\left(-,-\right)$ relative to which $S_\pm$ are lagrangian subspaces.  This means that the symplectic structure sets up an isomorphism $S_- \cong S_+^*$ of $\Spin(1,9)$ representations.  The representation $V$ is the real ten-dimensional vector representation of $\SO(1,9)$ with an invariant lorentzian inner product $\left<-,-\right>$.  The squaring map $\varphi: S_+ \times S_+ \to V$ is defined as in the case of eleven-dimensional supergravity as the transpose of the Clifford product $V \times S_+ \to S_-$ relative to the two inner products on $S$ and $V$; that is, for all $\varepsilon_{1,2} \in S_+$ and $v \in V$, $\varphi(\varepsilon_1,\varepsilon_2) \in V$ is defined by
\begin{equation}
 \left<v, \varphi(\varepsilon_1,\varepsilon_2)\right> = \left(\varepsilon_1, v \cdot \varepsilon_2\right)~.
\end{equation}
As in the case of eleven-dimensional supergravity, it again follows that for every nonzero $\varepsilon \in S_+$, $\varphi(\varepsilon,\varepsilon) \in V$ is nonzero and is not spacelike.

The values at $p$ of the Killing spinors define a subspace $W \subset S_+$.  We wish to show that if $\dim W > 8$, then the restriction $\varphi|_W$ of $\varphi$ to $W$ is surjective onto $V$.  Let $\dim W > 8$.  A vector $v \in V$ is perpendicular to the image of $\varphi|_W$ if and only if for all $\varepsilon_{1,2} \in W$,
\begin{equation}
  0 = \left<v, \varphi(\varepsilon_1,\varepsilon_2)\right> = \left(\varepsilon_1, v \cdot \varepsilon_2\right)~.
\end{equation}
In other words, the Clifford product with $v$ sends $W$ to its annihilator
\begin{equation}
  W^0 = \left\{\chi \in S_- \middle | \left(\chi, \varepsilon\right) = 0 ~ \forall \varepsilon \in W\right\}
\end{equation}
in $S_-$.  Since $\dim W + \dim W^0 = 16$, $\dim W > 8$ implies that $\dim W > \dim W^0$ and hence the Clifford product with $v$ has nontrivial kernel, again purely by dimensional reasons.  However the Clifford relation $v^2 = -|v|^2 \1$ says that $v$ is null.  In other words, the perpendicular complement of the image of $\varphi|_W$ is a totally null subspace of $V$, hence at most one-dimensional.  We wish to show that it cannot be one-dimensional, so let us assume for a contradiction that it is.  Again we may choose a Witt basis $(e_+,e_-,e_i)$ for $V$ such that the image of $\varphi|_W$ is the null line spanned by $e_+$.  As in the case of eleven-dimensional supergravity, we observe that for every $\varepsilon \in W$, $\varphi(\varepsilon,\varepsilon)$ is a non-spacelike vector perpendicular to $e_+$, whence it has to be collinear with $e_+$, so that $\varphi(\varepsilon,\varepsilon) = \lambda(\varepsilon) e_+$ for some function $\lambda: W \to \RR$.  Again using polarisation we deduce that for all $\varepsilon_{1,2} \in W$, $\varphi(\varepsilon_1,\varepsilon_2)$ lies in the line spanned by $e_+$, whence the image of $\varphi|_W$ is one-dimensional, contradicting the fact that its codimension is equal to 1.  Therefore, we conclude that $\varphi|_W$ is surjective, as desired.

\subsection{Type IIB supergravity}
\label{sec:type-iib-supergr}

In type IIB supergravity, the relevant spinor representation is $\SS_+ = S_+ \oplus S_+$, consisting of two copies of the positive-chirality spinor representation of $\Spin(1,9)$.  Letting $\SS_- = S_- \oplus S_-$, we have that $\SS = \SS_+ \oplus \SS_-$ consists of two copies of the unique irreducible module $S=S_+ \oplus S_-$ of $\Cl(1,9)$.  On $\SS$ we have an invariant symplectic inner product which is given by the diagonal extension of the one on $S$ discussed in the previous section.  In other words, if we let $\varepsilon_{1,2} \in \SS$, then their symplectic inner product is given by
\begin{equation}
  \left(\varepsilon_1,\varepsilon_2\right) = \sum_{A=1}^2  \left(\varepsilon^A_1,\varepsilon^A_2\right)~,
\end{equation}
where $\varepsilon_1 =\begin{pmatrix} \varepsilon_1^1 \\ \varepsilon_1^2\end{pmatrix}$, et cetera.  Relative to this symplectic inner product, $\SS_\pm$ are lagrangian subspaces.

Here $V$ is the ten-dimensional vector representation of $\Spin(1,9)$ as in the case of type I/heterotic supergravity.  Its invariant lorentzian inner product is denoted as before by $\left<-,-\right>$.
The squaring map $\varphi: \SS_+ \times \SS_+ \to V$ is defined again in such a way that for all $v \in V$ and $\varepsilon_{1,2} \in \SS_+$,
\begin{equation}
  \left<v,\varphi(\varepsilon_1,\varepsilon_2)\right> = \left(\varepsilon_1,v \cdot \varepsilon_2\right)~,
\end{equation}
or in more traditional notation and again relative to pseudo-orthonormal basis $(e_\mu)$ for $V$,
\begin{equation}
  \varphi(\varepsilon_1,\varepsilon_2) = \sum_A \bar\varepsilon_1^A \Gamma^\mu \varepsilon_2^A e_\mu
\end{equation}
where the sum over $\mu$ is implicit.

It follows as before that if $\varepsilon \in \SS_+$ is nonzero, then the vector $v$ with components $v_\mu = \sum_A \bar\varepsilon^A \Gamma_\mu \varepsilon^A$ is nonzero and is not spacelike, since
\begin{equation}
  v_0 = \sum_A \bar\varepsilon^A \Gamma_0 \varepsilon^A = \sum_A (\varepsilon^A)^\dagger \Gamma_0 \Gamma_0 \varepsilon^A = - \sum_A (\varepsilon^A)^\dagger \varepsilon^A < 0~.
\end{equation}

Now let $W \subset \SS_+$ denote the subspace defined by the values at $p$ of the Killing spinors and let $\varphi|_W$ denote the restriction of $\varphi$ to $W$.  A vector $v \in V$ is perpendicular to the image of $\varphi|_W$ if and only if Clifford product with $v$ maps $W$ to its annihilator $W^0 \in \SS_-$.  If $\dim W > 16$, $\dim W^0 < 16$, whence Clifford product with $v$ has nonzero kernel and hence by the Clifford relation $v$ is null.  Therefore $(\im \varphi|_W)^\perp$ is a totally null subspace of $V$ and hence it must be at most one-dimensional.  Assuming for a contradiction that it is one-dimensional and choosing a Witt basis $(e_+,e_-,e_i)$ for $V$ so that $(\im\varphi|_W)^\perp = \RR e_+$, we again conclude by polarisation that $\im\varphi|_W = \RR e_+$, which contradicts the codimension of $\im \varphi|_W$ being $1$.  Therefore we can conclude that $\varphi|_W$ is surjective as desired.

\section{Conclusion}
\label{sec:conclusion}

We have given uniform and elementary proofs of the strong homogeneity conjecture for the ten- and eleven-dimensional Poincaré supergravity theories.  This result simplifies the classification efforts of backgrounds preserving more than one half of the supersymmetry, by restricting the search to homogeneous backgrounds.  Of course, classifying homogeneous lorentzian manifolds is not an easy matter, but one can go some way towards a classification with some further restrictions (e.g., semisimplicity of $\fk_0$).

We believe that the strong homogeneity conjecture is true in more generality and an effort is currently underway to study its validity in other supergravity theories.  It may also be the case that one can generalise these results to other geometric situations, strengthening the work in \cite{ACKilling}.

\section*{Acknowledgments}

This work was supported in part by the grant ST/J000329/1 ``Particle Theory at the Tait Institute'' from the UK Science and Technology Facilities Council.  In addition, NH is supported by an EPSRC studentship.

\bibliographystyle{utphys}
\bibliography{AdS,AdS3,ESYM,Sugra,Geometry,CaliGeo}

\end{document}